\documentclass{jpp}
\usepackage{graphicx}
\usepackage{color}
\usepackage{amssymb}
\usepackage{amsmath}
\usepackage{tabularx}
\usepackage{bm}
\usepackage{ltxtable}
\usepackage{natbib} 
\usepackage{longtable}
\usepackage{float}

\newcommand{\bB}{{\bf B}}
\newcommand{\bU}{{\bf U}}

\newcommand{\be}{{\bf e}}
\newcommand{\bE}{{\bf E}}

\newcommand{\bH}{{\bf H}}
\newcommand{\bJ}{{\bf J}}

\newcommand{\bq}{\textit{\textbf{q}}}

\newcommand{\ii}{\mathrm{i}}

\shorttitle{Dynamo action with anisotropic conductivity and permeability}
\shortauthor{F. Plunian, T. Alboussi\`ere}

\title{Axisymmetric dynamo action\\ produced by differential rotation, \\with anisotropic electrical conductivity \\and anisotropic magnetic permeability}

\author{Franck Plunian\aff{1}
  \corresp{\email{Franck.Plunian@univ-grenoble-alpes.fr}},
 \and 
 Thierry Alboussi\`ere\aff{2}}

\affiliation{\aff{1}Universit\'e Grenoble Alpes, Universit\'e Savoie Mont Blanc, CNRS, IRD, IFSTTAR, ISTerre, 38000 Grenoble, France
\aff{2}Univ. Lyon, Univ. Lyon 1, ENSL, CNRS, LGL-TPE, F-69622, Villeurbanne, France}

\begin{document}

\maketitle

\begin{abstract}
The effect on dynamo action of an anisotropic electrical conductivity conjugated to an anisotropic magnetic permeability is considered.
Not only is the dynamo fully axisymmetric, but it requires only a simple differential rotation, which twice challenges the well-established dynamo theory.
Stability analysis is conducted entirely analytically, leading to an explicit expression of the dynamo threshold.  
The results show a competition between the anisotropy of electrical conductivity and that of magnetic permeability, the dynamo effect becoming impossible if the two anisotropies are identical.
For isotropic electrical conductivity, Cowling's neutral point argument does imply the absence of an azimuthal component of current density, but does not prevent the dynamo effect as long as the magnetic permeability is anisotropic. 
\end{abstract}


\section{Introduction}
%%%%%%%%%%
The dynamo effect is a magnetic instability produced by the displacement of an electrically conducting medium, without the aid of a magnet or a remanent magnetic field.
A part of the kinetic energy of the moving medium is thus transferred into magnetic energy.
This process is the most likely candidate to explain the ubiquity of magnetic fields observed in astrophysical objects \citep{Rincon2019}.  
The increasing resolution of numerical simulations of dynamo equations makes it possible to reproduce, ever better, the magnetic features measured in natural objects \citep{Schaeffer2017}. 
Experiments have also successfully demonstrated the possibility of dynamo action \citep{Gailitis2001, Stieglitz2001, Monchaux2007}, although it remains difficult to replicate in the laboratory processes occurring on a geophysical or astrophysical scale \citep{Alboussiere2011,Tigrine2019}. This is why, since the very first dynamo experiments \citep{Lowes1963,Lowes1968}, the use of materials with the highest product of electrical conductivity and magnetic permeability, has been favoured. In most cases it led to the choice of solid iron alloys (high permeability) or liquid sodium (high conductivity) for the moving parts. In the VKS experiment, in which liquid sodium was driven by impellers, the high magnetic permeability of the impellers was revealed to be crucial to achieve a dynamo effect \citep{Miralles2013, Kreuzahler2017, Nore2018}. The choice of materials for the static parts, like the walls of the container, has also been proved to be crucial in relation to the electromagnetic boundary conditions \citep{Avalos2003,Avalos2005}, leading for example to the use of copper walls \citep{Monchaux2007}. 

The role in the dynamo effect of an anisotropic electrical conductivity has been studied for different geometries: Cartesian \citep{Ruderman1984, Alboussiere2020}, cylindrical \citep{Plunian2020} and toroidal \citep{Lortz1989}. 
Although at first glance it is difficult to imagine such an anisotropic electrical conductivity in natural objects, it is far from impossible.
For example, in a plasma subjected to a magnetic field, it is known that the electrical conductivity in the direction parallel to the magnetic field is twice that in the direction perpendicular to the magnetic field \citep{Braginskii1965}. In the case of the Earth, seismic observations provide strong evidence that the elastic response of the solid inner core is anisotropic. This is most likely due to the alignment of hexagonal close-packed iron crystals, occurring during the solidification of the inner core \citep{Deuss2014}. Incidentally it has been shown that, in hexagonal close-packed iron, the thermal conductivity and the electrical conductivity, which are directly related, are anisotropic \citep{Ohta2018}. Eventually, this suggests that the electrical conductivity of the inner core is anisotropic.
Finally, in a turbulent electrically conducting fluid, the interaction between the small scales of the velocity field and those of the magnetic field can generate a large-scale magnetic field by dynamo action. Such a process can be modeled using the so-called mean-field approach \citep{Krause1980}, possibly leading to a large-scale anisotropic electrical conductivity \citep{brandenburg2018}.

From a theoretical point of view, an interesting consequence of considering an anisotropic conductivity is the possibility of obtaining an axisymmetric dynamo effect \citep{Plunian2020}, allowing one to bypass the well-known Cowling's antidynamo theorem \citep{Cowling1934,Kaiser2014}. Indeed, if it is anisotropic, then the electrical conductivity becomes a tensor, instead of being a scalar, which defeats Cowling's neutral point argument. 
In addition, the use of an anisotropic conductivity leads to dynamo action for a motion as simple as shear \citep{Ruderman1984,Alboussiere2020,Plunian2020}, which  is otherwise impossible to achieve. 

Here we investigate the role of an anisotropic electrical conductivity conjugated to an anisotropic magnetic permeability.
An anisotropic magnetic permeability is not expected in natural objects. However, as explained above, this may be of interest for dynamo experiments, in order to reduce the dynamo threshold. 
In contrast to previous dynamo studies \citep{Busse1992,Kaiser1999}, here the electrical conductivity and magnetic permeability do not depend on time or space coordinates. In addition, they are stationary and axisymmetric.

\section{Conductivity and permeability anisotropy}
%%%%%%%%%%%%%%%%%%
We consider a material such that the electrical conductivity and magnetic permeability are denoted $\sigma^{\parallel}$ and $\mu^{\parallel}$ in a given direction $\bq$, and
$\sigma^{\perp}$ and $\mu^{\perp}$ in the directions perpendicular to $\bq$. 

Writing Ohm's law, $\bJ=\sigma^{\parallel} \bE$ in the direction of $\bq$ and $\bJ=\sigma^{\perp} \bE$ in the directions perpendicular to $\bq$, leads to the following conductivity tensor: 
\begin{equation}
\lbrack\sigma_{ij}\rbrack=\sigma^{\perp} \delta_{ij} + (\sigma^{\parallel}-\sigma^{\perp})q_i q_j.
\label{eq:conductivity tensor}
\end{equation}
Inverting (\ref{eq:conductivity tensor}) leads to the resistivity tensor: \citep{Ruderman1984}
\begin{equation}
\lbrack\sigma_{ij}\rbrack^{-1}=\frac{1}{\sigma^{\perp}} \delta_{ij} + (\frac{1}{\sigma^{\parallel}}-\frac{1}{\sigma^{\perp}})q_i q_j.
\label{eq:resistivity tensor}
\end{equation}
Similarly, a magnetic permeability tensor can be defined as
\begin{equation}
\lbrack\mu_{ij}\rbrack=\mu^{\perp} \delta_{ij} + (\mu^{\parallel}-\mu^{\perp})q_i q_j,
\label{eq:permeability tensor}
\end{equation}
with the inverse tensor
\begin{equation}
\lbrack\mu_{ij}\rbrack^{-1}=\frac{1}{\mu^{\perp}} \delta_{ij} + (\frac{1}{\mu^{\parallel}}-\frac{1}{\mu^{\perp}})q_i q_j.
\label{eq:inv permeability tensor}
\end{equation}

We choose $\bq$ as a unit vector in the horizontal plane:
 \begin{equation}
 \bq=c \; \be_r+s \; \be_{\theta},
 \end{equation}
 where $(\be_r,\be_{\theta},\be_z)$ is a cylindrical coordinate system, with $c=\cos\alpha$ and $s=\sin\alpha$, $\alpha$ being a prescribed angle. 

\begin{figure}
\centering
\includegraphics[scale=0.4]{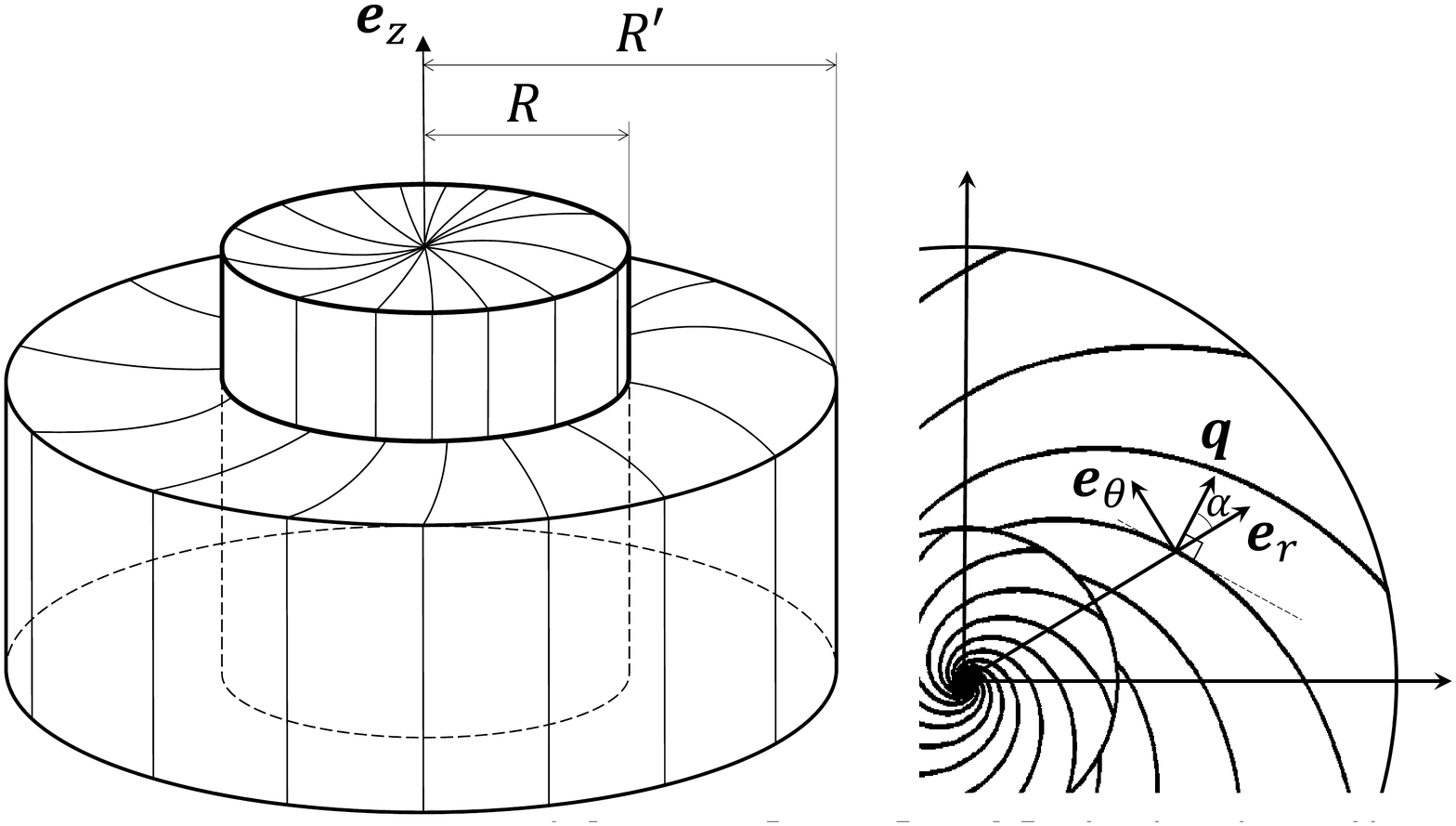}
\caption{Left: The inner-cylinder of radius $R$ rotates as a solid-body within an outer cylinder at rest. The radius $R'$ of the outer cylinder is taken as
 infinite. 
Right: The curved lines are perpendicular to $\bq$ and describe logarithmic spirals. They correspond to the directions along which 
$\sigma=\sigma^{\perp}$ and $\mu=\mu^{\perp}$.}
\label{fig:spirales}
\end{figure}

  In figure ~\ref{fig:spirales} the curved lines are perpendicular to $\bq$ and describe logarithmic spirals. They correspond to the directions along which 
$\sigma=\sigma^{\perp}$ and $\mu=\mu^{\perp}$.
We consider the solid-body rotation $\bU$
of a cylinder of radius $R$ embedded in an infinite medium at rest. Both regions are made of the same material, with therefore identical conductivity tensors and identical permeability tensors. 

\section{Induction equation}
%%%%%%%%%%%%%
%%%%%%%%%%%%%

In the magnetohydrodynamic approximation, the Maxwell equations and Ohm's law take the form
\begin{eqnarray}
\bH=\lbrack\mu_{ij}\rbrack^{-1}\bB,\label{eq:Maxwell1}\\
 \bJ=\nabla \times \bH, \label{eq:Ampere}\\
 \partial_t \bB = - \nabla \times \bE\label{eq:Maxwell3}\\
\nabla \cdot \bB = 0, \label{eq:Maxwell4}\\
\bJ=\lbrack\sigma_{ij}\rbrack(\bE + \bU \times \bB),\label{eq:Ohm}
\end{eqnarray}
where $\bH$, $\bB$, $\bJ$, $\bE$ and $\bU$ are the magnetic field, the induction field, the current density, the electric field and the velocity field.
The induction equation then takes the form
\begin{equation}
\partial_t \bB = \nabla \times (\bU \times \bB) - \nabla \times \left( \lbrack\sigma_{ij}\rbrack^{-1} \nabla \times \left( \lbrack\mu_{ij}\rbrack^{-1}\bB \right) \right).
\label{eq:induction equation}
\end{equation}

Renormalizing the distance, electrical conductivity, magnetic permeability and time by respectively $R, \mu^{\perp}, \sigma^{\perp}$ and $\mu^{\perp}\sigma^{\perp}R^2$, the dimensionless form of the induction equation is identical to (\ref{eq:induction equation}), but
with 
\begin{eqnarray}
\lbrack\sigma_{ij}\rbrack^{-1} &=&  \delta_{ij}+\sigma q_i q_j, \;\;\; \sigma=\frac{\sigma^{\perp}}{\sigma^{\parallel}}-1, \label{eq:dimensionless conductivity}\\
\lbrack\mu_{ij}\rbrack^{-1} &=&  \delta_{ij}+\mu q_i q_j, \;\;\; \mu=\frac{\mu^{\perp}}{\mu^{\parallel}}-1 \label{eq:dimensionless permeability}
\end{eqnarray}
and
\begin{equation}
\bU=
\left\{
\begin{split}
r\Omega \be_{\theta}&,&r< 1 \\
0&,&r>1
\end{split}\;\;\;,
\right. \label{eq:velocity}
\end{equation}
where $\Omega$ is the dimensionless angular velocity of the inner cylinder. We note that $(\sigma, \mu) \in [-1, +\infty[^2$, with $\sigma=0$ and $\mu=0$ corresponding to respectively isotropic conductivity and isotropic permeability.

Provided the velocity is stationary and $z$-independent, an axisymmetric magnetic induction can be searched in the form
\begin{equation}
\bB(r,z,t)= \tilde{B}\be_{\theta} + \nabla \times \left( \tilde{A}\be_{\theta}\right),
\end{equation}
with $(\tilde{A}, \tilde{B})=(A(r), B(r))\exp(\gamma t + \ii kz)$, where $\gamma$ is the instability growth rate and $k$ the vertical wavenumber of the corresponding eigenmode, and
where $A$ and $B$ depend only on the radial coordinate $r$. 
Thus the magnetic induction takes the form
\begin{equation}
\bB=\left(-\ii k A, B, \frac{1}{r}\partial_r(rA) \right)\exp(\gamma t + \ii kz),
\label{eq:magnetic induction}
\end{equation}
with dynamo action corresponding to $\Re\{\gamma\}>0$.

From (\ref{eq:velocity}) and (\ref{eq:magnetic induction}),  it can be shown that $\nabla\times(\bU\times\bB)=0$ in each region $r<1$ and $r>1$ (see appendix \ref{sec: curl of u times B}).
Replacing (\ref{eq:dimensionless conductivity}), (\ref{eq:dimensionless permeability}) and (\ref{eq:magnetic induction}) into the induction equation (\ref{eq:induction equation}) leads to
\begin{eqnarray}
\gamma A + (1+\sigma s^2)D_k(A)  + \mu c^2 k^2 A &=&
\ii cs k (\sigma-\mu) B, \label{eq:gamma A}\\
\gamma B + (1+\mu s^2)D_k(B) + \sigma c^2 k^2 B&=&
- \ii csk (\sigma-\mu) D_k(A)    \label{eq:gamma B},
\end{eqnarray}
where $D_{\nu}(X)=\nu^2X-\partial_r\left(\frac{1}{r}\partial_r(rX)\right)$. 
The derivation of (\ref{eq:gamma A}) and (\ref{eq:gamma B}) is given in appendix \ref{sec: eq of A and B}.
For $\sigma=\mu=0$, corresponding to isotropy of both conductivity and permeability, (\ref{eq:gamma A}) and (\ref{eq:gamma B}) are diffusion equations, leading to a free decaying solution (no dynamo action).
For an isotropic permeability, $\mu=0$, (\ref{eq:gamma A}) and (\ref{eq:gamma B}) are identical to the equations derived in \citet{Plunian2020}.

\section{Dynamo threshold}
%%%%%%
%%%%%%%%%%%% 
\subsection{General form of the solutions}
%%%%%%%%%%%%%%%%%%%%%
Looking for non-oscillating solutions, the dynamo threshold then corresponds to $\gamma=0$. 
Thus, taking $\gamma=0$ in (\ref{eq:gamma A}) and (\ref{eq:gamma B}), it can be shown (Appendix \ref{sec:derivation of DoD}) that 
\begin{equation}
\left(D_{k_{\mu}}\circ D_{k_{\sigma}}\right)(A)=\left(D_{k_{\sigma}}\circ D_{k_{\mu}}\right)(B) = 0, \label{eq:differential equations}
\end{equation}
where 
\begin{equation}
k_{\sigma}=k\left(\frac{1+\sigma}{1+\sigma s^2}\right)^{1/2}, \;\;\;\;\; k_{\mu}=k\left(\frac{1+\mu}{1+\mu s^2}\right)^{1/2}.
\label{eq:ksigma kmu}
\end{equation}
We note that the two operators $D_{k_{\sigma}}$ and $D_{k_{\mu}}$ are commutative. 
Therefore in (\ref{eq:differential equations}) we can apply the two operators in the order we want, $D_{k_{\mu}}\circ D_{k_{\sigma}}$  or $D_{k_{\sigma}}\circ D_{k_{\mu}}$, to both $A$ and $B$.
The set of functions $X(r)$, satisfying the fourth-order differential equation 
$\left(D_{k_{\mu}}\circ D_{k_{\sigma}}\right)(X)=0$, is a vector space of dimension 4.  
Now, we know that, whatever $\nu$, the solutions of $D_{\nu}(X)=0$ are a linear combination of $I_1(\nu r)$ and $K_1(\nu r)$, where $I_1$ and $K_1$ are modified Bessel functions of first and second kind, of order 1.
Therefore the solutions of (\ref{eq:differential equations})
are a linear combination of $I_1(k_{\sigma} r)$, $K_1(k_{\sigma} r)$, $I_1(k_{\mu} r)$ and $K_1(k_{\mu} r)$.

Looking for $A$ in the form
\begin{equation}
A= \alpha_{\sigma} I_1(k_{\sigma} r)+ \beta_{\sigma} K_1(k_{\sigma} r) + \alpha_{\mu} I_1(k_{\mu} r) + \beta_{\mu} K_1(k_{\mu} r),
\end{equation}
and specifying that $A$ must be finite at $r=0$ and that $\lim\limits_{r \to \infty}A=0$, leads to
\begin{eqnarray}
A=&&
\left\{
\begin{split}
r< 1,&\;\;\;\;\;\alpha_{\sigma} I_1(k_{\sigma} r)+ \alpha_{\mu} I_1(k_{\mu} r)&  \\
r> 1,&\;\;\;\;\;\beta_{\sigma} K_1(k_{\sigma} r) + \beta_{\mu} K_1(k_{\mu} r)&
\end{split}
\right. , \label{eq:Aparameters}
\end{eqnarray}
where $\alpha_{\sigma}$, $\beta_{\sigma}$, $\alpha_{\mu}$ and $\beta_{\mu}$ are free parameters that will be constrained by additional boundary conditions at $r=1$.
Replacing (\ref{eq:Aparameters}) in (\ref{eq:gamma A}) for $\gamma=0$ leads to the following expression for $B$

\begin{eqnarray}
B=&&
\left\{
\begin{split}
r< 1,&\;\;\;\frac{\ii ck}{s} \left(\alpha_{\sigma} I_1(k_{\sigma} r)+ \frac{\mu s^2}{1+\mu s^2}\alpha_{\mu} I_1(k_{\mu} r)\right) &  \\
r> 1,&\;\;\;\frac{\ii ck}{s} \left(\beta_{\sigma} K_1(k_{\sigma} r) + \frac{\mu s^2}{1+\mu s^2}\beta_{\mu} K_1(k_{\mu} r) \right)&
\end{split}
\right. , \label{eq:Bparameters}
\end{eqnarray}
the derivation of which being given in Appendix \ref{sec:Derivation de B}.
%%%%%%%%%%%%%%
%%%%%%%%%%%%%%
%%%%%%%%%%%%%%

\subsection{Boundary conditions at $r=1$}
\label{sec:boundary conditions}
%%%%%%%%%%%%%%%
From the Maxwell equations and Green-Ostrogradski and Stokes theorems, the radial component of $\bB$ and the tangential components of $\bH=\lbrack\mu_{ij}\rbrack^{-1}\bB$ must be continuous at $r=1$.
Taking the expression of $\bB$ and $\bH$ given in (\ref{eq:magnetic induction}) and (\ref{eq:magnetic field}), these continuity conditions 
can be written in terms of $A$ and $B$ as
\begin{eqnarray}
A(r=1^-)&=&A(r=1^+), \label{eq:contA}\\
B(r=1^-)&=&B(r=1^+),\label{eq:contB}\\
\partial_r A(r=1^-)&=&\partial_r A(r=1^+).\label{eq:contdA}
\end{eqnarray}

Taking $A$ and $B$ given in (\ref{eq:Aparameters}) and  (\ref{eq:Bparameters}) and replacing them in (\ref{eq:contA}), (\ref{eq:contB}) and (\ref{eq:contdA})
leads to (Appendix \ref{sec:derivation of BC})
\begin{eqnarray}
\alpha_{\sigma} I_1(k_{\sigma})-\beta_{\sigma} K_1(k_{\sigma})&=&0\label{eq:BC1}\\
\alpha_{\mu} I_1(k_{\mu})-\beta_{\mu} K_1(k_{\mu})&=&0\label{eq:BC2}\\
\alpha_{\sigma}k_{\sigma} I_0(k_{\sigma})+ \alpha_{\mu}k_{\mu} I_0(k_{\mu})+
\beta_{\sigma}k_{\sigma} K_0(k_{\sigma})+ \beta_{\mu}k_{\mu} K_0(k_{\mu})&=&0, \label{eq:contdA2}
\end{eqnarray}
where $I_0$ and $K_0$ are modified Bessel functions of first and second kind, of order 0.
It is convenient to introduce the parameters $\lambda_{\sigma}=\alpha_{\sigma} I_1(k_{\sigma})\ii k / s$ and $\lambda_{\mu}=\alpha_{\mu} I_1(k_{\mu})\ii k / s$. Then, using (\ref{eq:BC1}) and (\ref{eq:BC2}), we can rewrite $A$ and $B$ in the following form:
\begin{eqnarray}
\ii kA=&&
\left\{
\begin{split}
r< 1,&\;\;\;\;\;s\left(\lambda_{\sigma} \frac{I_1(k_{\sigma} r)}{I_1(k_{\sigma})}+ \lambda_{\mu} \frac{I_1(k_{\mu} r)}{I_1(k_{\mu})}\right)&  \\
r> 1,&\;\;\;\;\;s\left(\lambda_{\sigma} \frac{K_1(k_{\sigma} r)}{K_1(k_{\sigma})}+ \lambda_{\mu} \frac{K_1(k_{\mu} r)}{K_1(k_{\mu})}\right)&
\end{split}
\right. , \label{eq:Aparameters2}
\end{eqnarray}

\begin{eqnarray}
B=&&
\left\{
\begin{split}
r< 1,&\;\;\;c \left(\lambda_{\sigma} \frac{I_1(k_{\sigma} r)}{I_1(k_{\sigma})}+ \frac{\mu s^2}{1+\mu s^2}\lambda_{\mu} \frac{I_1(k_{\mu} r)}{I_1(k_{\mu})}\right) &  \\
r> 1,&\;\;\;c \left(\lambda_{\sigma} \frac{K_1(k_{\sigma} r)}{K_1(k_{\sigma})}+ \frac{\mu s^2}{1+\mu s^2}\lambda_{\mu} \frac{K_1(k_{\mu} r)}{K_1(k_{\mu})} \right)&
\end{split}
\right. . \label{eq:Bparameters2}
\end{eqnarray}

\begin{figure}
\centering
\includegraphics[scale=0.4]{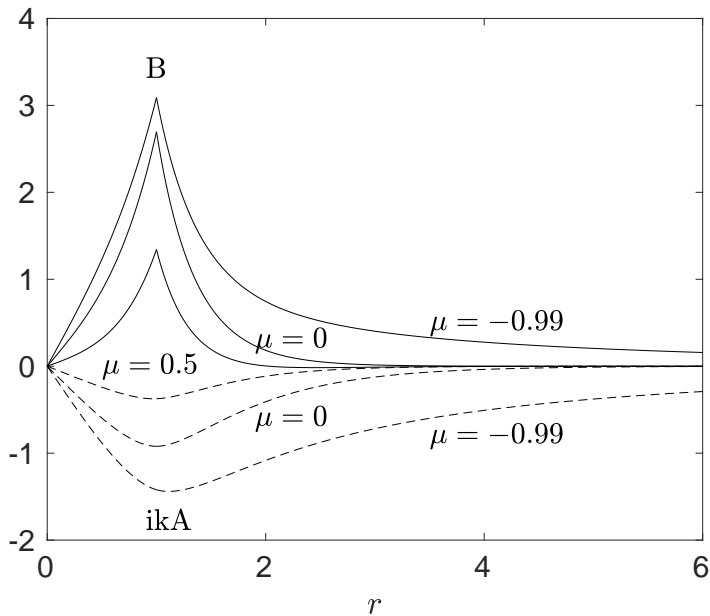}
\caption{Eigenmodes $\ii kA$ (dashed lines) and $B$ (solid lines) versus $r$, for $\sigma=10^6, k=1.1, \alpha=0.16\pi$, $\lambda_{\sigma}=\Gamma(k_{\mu})$, $\lambda_\mu=-\Gamma(k_{\sigma})$ and for $\mu \in\{-0.99, 0, 0.5\}$.}
\label{fig:eigenmodes}
\end{figure}

The continuity of  $\partial_rA$ at $r=1$, given by (\ref{eq:contdA2}), then leads to the following identity between $\lambda_{\sigma}$ and $\lambda_{\mu}$
 \begin{equation}
 \lambda_{\sigma} \Gamma(k_{\sigma}) + \lambda_\mu \Gamma(k_{\mu})=0,
 \label{eq:A'cont}
 \end{equation}
 with
 \begin{equation}
 \Gamma(x)=x\left(\frac{I_0(x)}{I_1(x)} + \frac{K_0(x)}{K_1(x)}\right) \equiv \left(I_1(x) K_1(x)\right)^{-1},
 \label{eq:Gamma}
 \end{equation}
the last equality coming from the Wronskian relation 
\begin{equation}
I_m(x)K_{m+1}(x)+I_{m+1}(x)K_m(x)=1/x. \label{eq:Wronskian}
\end{equation}
In figure \ref{fig:eigenmodes} the eigenmodes $\ii kA$ and $B$ are plotted versus $r$ for $\lambda_{\sigma}=\Gamma(k_{\mu})$ and $\lambda_\mu=-\Gamma(k_{\sigma})$ such that (\ref{eq:A'cont}) is satisfied. 

Finally, the 
tangential components $\bE_{\theta}$ and $\bE_z$ of the electric field 
\begin{equation}
\bE=- \bU \times \bB + \lbrack\sigma_{ij}\rbrack^{-1}\bJ \label{eq:EE}
\end{equation}
 have to be continuous at $r=1$. 
The expression of the current density $\bJ$, which is derived in Appendix \ref{sec:derivation of j}, is given by
\begin{eqnarray}
\bJ_r=&&-\ii c k \lambda_{\sigma}
\left\{
\begin{split}
r< 1,&\;\;\;I_1(k_{\sigma} r)/I_1(k_{\sigma})   \\
r> 1,&\;\;\;K_1(k_{\sigma} r)/K_1(k_{\sigma})
\end{split}
\right.  ,\label{eq:jr}
\end{eqnarray}
\begin{eqnarray}
\bJ_{\theta}=&&- \frac{\sigma s c}{1+\sigma s^2} \bJ_r,\label{eq:jtheta}
\end{eqnarray}
\begin{eqnarray}
\bJ_z=&&c k_{\sigma} \lambda_{\sigma}
\left\{
\begin{split}
r< 1,&\;\;\;I_0(k_{\sigma} r)/I_1(k_{\sigma})   \\
r> 1,&\;\;\;-K_0(k_{\sigma} r)/K_1(k_{\sigma})
\end{split}
\right. , \label{eq:jz}
\end{eqnarray}
where the coefficient $\exp(\ii k z)$ has been dropped for convenience.

From (\ref{eq:dimensionless conductivity}), (\ref{eq:velocity}) and (\ref{eq:jtheta}), we find that $\bE_{\theta}=0$, which is in agreement with axisymmetric solutions.
Indeed, Maxwell equation (\ref{eq:Maxwell3}) taken at the threshold implies $\nabla \times \bE = 0$. Applying the Stokes theorem to the integral of $\nabla \times \bE$ on a disc of radius $r$, and assuming axisymmetry, then leads to $\bE_{\theta}(r)=0$.

The continuity of  $\bE_z$ implies the following identity:
\begin{equation}
(\bJ_z + r \Omega \bB_r)(r=1^-) = \bJ_z(r=1^+).
\label{eq:Ezcont}
\end{equation}
Replacing (\ref{eq:Aparameters2}) and (\ref{eq:jz}) in (\ref{eq:Ezcont}), and using (\ref{eq:A'cont}), leads to the dynamo threshold
\begin{equation}
\Omega^c = \frac{c}{s}\left(I_1(k_{\sigma}) K_1(k_{\sigma})-I_1(k_{\mu})K_1(k_{\mu})\right)^{-1}.
\label{eq:Omegac}
 \end{equation}
\begin{figure}
\centering
\includegraphics[scale=0.4]{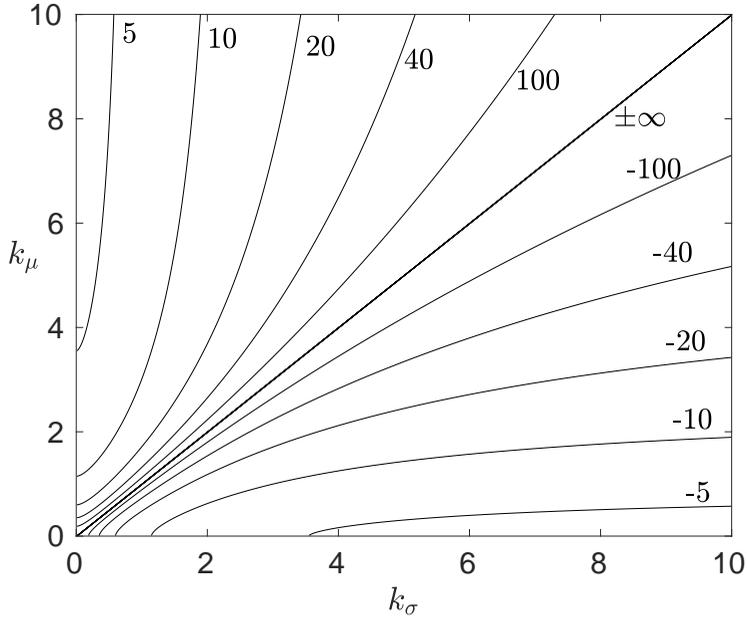}
\caption{Isovalues of $\Omega^c\in\{\pm5, \pm10, \pm20, \pm40, \pm100\}$ in the ($k_{\sigma},k_{\mu}$) map, for $\alpha=0.16\pi$. The diagonal $k_{\sigma}=k_{\mu}$ corresponds to $\Omega^c\rightarrow \pm \infty$.}
\label{fig:ksigma_kmu}
\end{figure}
\begin{figure}
\centering
\includegraphics[scale=0.4]{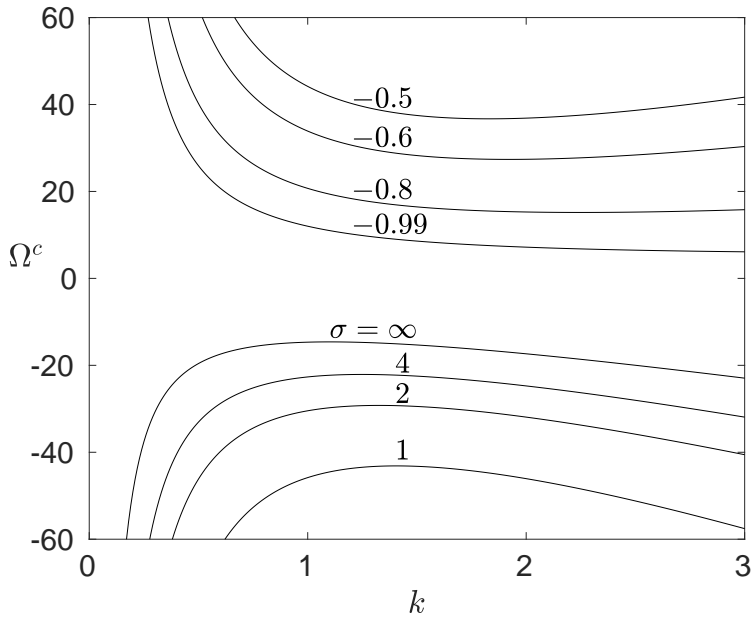}
\caption{Curves of the dynamo threshold $\Omega^c$ versus $k$, for $\mu=0$, $\alpha=0.16\pi$ and $\sigma\in\{-0.99, -0.8,-0.6, -0.5, 1, 2, 4, +\infty\}$.}
\label{fig:Omegac_k_mu_sigma0}
\end{figure}
\begin{figure}
\centering
\includegraphics[scale=0.4]{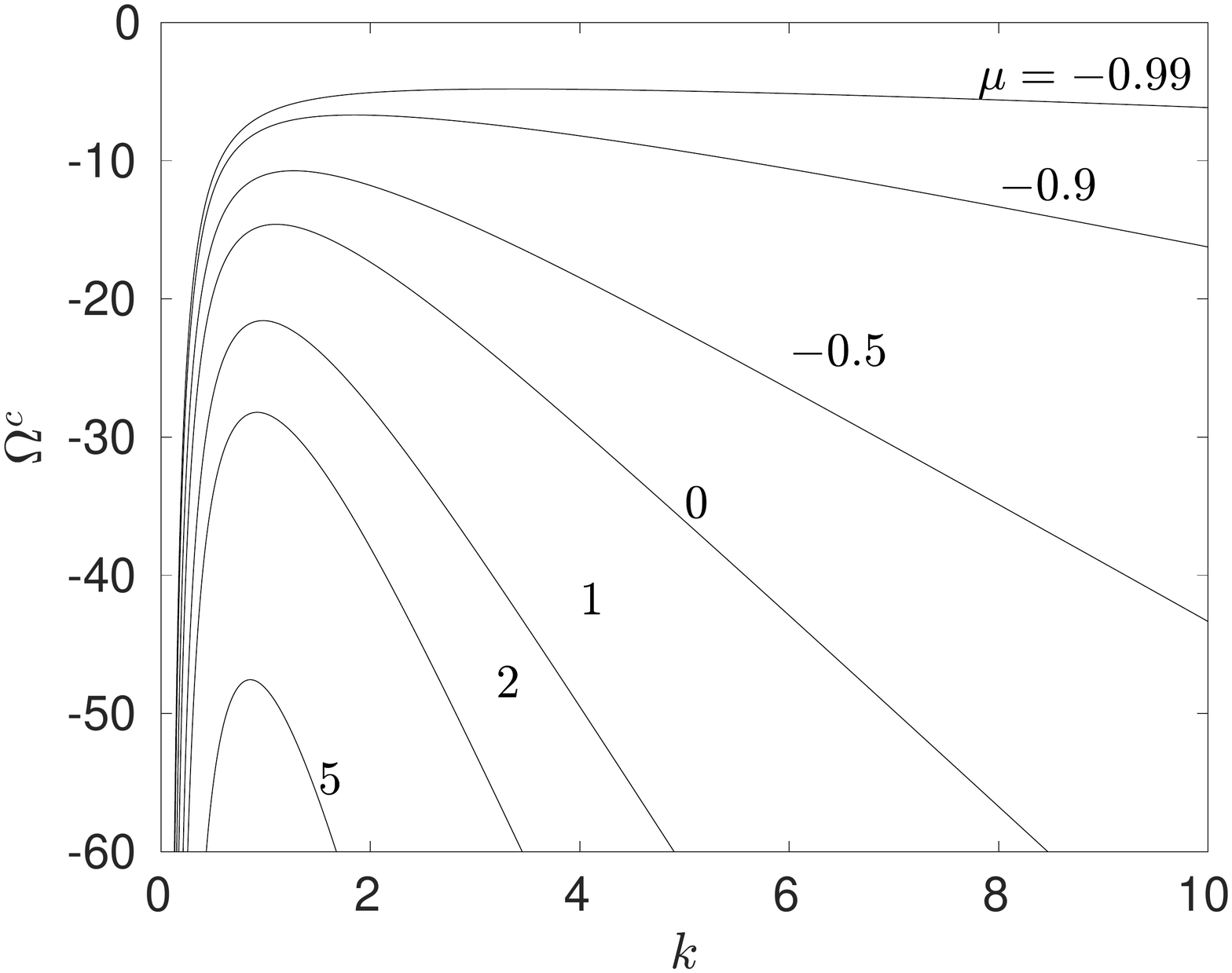}
\caption{Curves of the dynamo threshold $\Omega^c$ versus $k$, for $\sigma=10^5$, $\alpha=0.16\pi$ and $\mu\in\{ -0.99, -0.9, -0.5, 0, 1, 2, 5\}$.}
\label{fig:marginal curves}
\end{figure}

\section{Analysis of the results}
%%%%%%%%%%%%%%%
\subsection{Dispersion relation}

A striking consequence of (\ref{eq:Omegac}) is that $\Omega^c(\sigma,\mu)$ is antisymmetric, satisfying
\begin{equation}
\Omega^c (\sigma,\mu) = - \Omega^c (\mu,\sigma). \label{eq:antisymmetry Omegac}
\end{equation}
In addition for identical anisotropies of conductivity and permeability, $\mu=\sigma$, the threshold is infinite, leading to the impossibility of an axisymmetric dynamo,
\begin{equation}
\lim\limits_{|\sigma-\mu|\rightarrow 0}|\Omega^c (\sigma,\mu)| \rightarrow +\infty. \label{eq:Omegac infinite}
\end{equation}
This is illustrated in figure \ref{fig:ksigma_kmu}, in which the curves of a few isovalues of $\Omega^c$ are plotted versus $k_{\sigma}$ and $k_{\mu}$.
In particular, having both $\sigma \gg 1$ and $\mu \gg 1$ is detrimental for dynamo action, as in this case, from (\ref{eq:ksigma kmu}), $k_{\sigma} \approx k_{\mu} \approx k/s$. 

The antisymmetry property (\ref{eq:antisymmetry Omegac}) of $\Omega^c(\sigma,\mu)$ can also be derived directly from the set of equations (\ref{eq:gamma A}-\ref{eq:gamma B}) taken for $\gamma=0$, the boundary conditions (\ref{eq:contA}-\ref{eq:contdA}) and (\ref{eq:Ezcont}), without deriving explicitly the expressions of $A$ and $B$. This is shown in Appendix \ref{sec:derivation of antisymmetry}. 

Alternatively, changing $\alpha$ to $-\alpha$ in (\ref{eq:Omegac}) also changes $\Omega^c$ to $-\Omega^c$. 
This can be also derived directly from (\ref{eq:gamma A}-\ref{eq:gamma B}), taken for $\gamma=0$, the boundary conditions (\ref{eq:contA}-\ref{eq:contdA}), and (\ref{eq:Ezcont}),
by changing $A$ to $-A$ (or $B$ to $-B$).

 We check that for an isotropic permeability, $\mu=0$, $\Omega^c$ is the same as that given in \citet{Plunian2020}. 
In figure \ref{fig:Omegac_k_mu_sigma0} the curves of the dynamo threshold $\Omega^c$ versus $k$ are plotted for $\mu=0$ (isotropic magnetic permeability), $\alpha=0.16\pi$ and different values of $\sigma$. The negative values of $\sigma$ correspond to an electrical conductivity that is the highest in the direction parallel to $\bq$.

In figure \ref{fig:marginal curves} the curves of the dynamo threshold $\Omega^c$ versus $k$ are plotted for $\sigma=10^5$, $\alpha=0.16\pi$ and different values of $\mu$. For $\mu=0$, the minimum value of $|\Omega^c|$ is obtained for $k=1.1$ and  $\alpha=0.16 \pi$, and is equal to $\min_{k,\alpha}|\Omega^c|=14.61$ \citep{Plunian2020}. For positive values of  $\mu$, $|\Omega^c|$ increases with $\mu$, showing the detrimental effect of having both a high $\sigma$ and a high $\mu$.
For negative values of $\mu$, $|\Omega^c|$ decreases with $|\mu|$, showing that the dynamo effect is favoured if the permeability is higher in the direction parallel to $\bq$.

For $s=0$ (radial $\bq$) or $c=0$ (azimuthal $\bq$), the dynamo is impossible. This is obvious for $s=0$ as the threshold given by (\ref{eq:Omegac}) is infinite. For $c=0$, (\ref{eq:Bparameters2}) implies that $B=0$. In addition, as $s^2=1$, (\ref{eq:ksigma kmu}) implies that $k_{\sigma}=k_{\mu}=k$. Then, from (\ref{eq:Aparameters2}) and (\ref{eq:A'cont}), we find that $A=0$.

\subsection{Current density}
%%%%%%%%%%%%%
Concerning the current density $\bJ$, given at the threshold by ($\ref{eq:jr}$-$\ref{eq:jz}$), we note that it only depends on $\sigma$, and not on $\mu$. 
In other words, taking an anisotropic magnetic permeability $\mu \neq 0$ does not change the geometry of the current density with respect to the isotropic case $\mu=0$.

For an isotropic conductivity $\sigma=0$, we find that $\bJ_{\theta}=0$. This corresponds to the neutral point argument of \citet{Cowling1934}, after which a toroidal current density cannot be produced if axisymmetry is assumed.
However, and although such a neutral point argument is satisfied for $\sigma=0$, this does not exclude the possibility of dynamo action for an anisotropic magnetic permeability $\mu \neq 0$.

From (\ref{eq:jtheta}), we note that the projection in the ($r$,$\theta$) plane of the current density $\bJ$ describes spiralling trajectories. In the limit $\sigma\rightarrow +\infty$, we find that $\bJ \cdot \bq=0$.

\subsection{Magnetic induction}
%%%%%%%%%%%%%%%%
From the expression of $\bB$ given in (\ref{eq:magnetic induction}), and applying (\ref{eq:Aparameters2}), (\ref{eq:Bparameters2}) and (\ref{eq:dAdr}), leads to the following expressions for the magnetic induction components
\begin{eqnarray}
\bB_r=&&-s
\left\{
\begin{split}
r< 1,&\;\;\;\lambda_{\sigma}\frac{I_1(k_{\sigma} r)}{I_1(k_{\sigma})}+\lambda_{\mu}\frac{I_1(k_{\mu} r)}{I_1(k_{\mu})}   \\
r> 1,&\;\;\;\lambda_{\sigma}\frac{K_1(k_{\sigma} r)}{K_1(k_{\sigma})}+\lambda_{\mu}\frac{K_1(k_{\mu} r)}{K_1(k_{\mu})}
\end{split}
\right.  ,\label{eq:Br}
\end{eqnarray}
\begin{eqnarray}
\bB_{\theta}=&&c
\left\{
\begin{split}
r< 1,&\;\;\;\lambda_{\sigma}\frac{I_1(k_{\sigma} r)}{I_1(k_{\sigma})}+\lambda_{\mu}\left(\frac{\mu s^2}{1+\mu s^2}\right)\frac{I_1(k_{\mu} r)}{I_1(k_{\mu})}   \\
r> 1,&\;\;\;\lambda_{\sigma}\frac{K_1(k_{\sigma} r)}{K_1(k_{\sigma})}+\lambda_{\mu}\left(\frac{\mu s^2}{1+\mu s^2}\right)\frac{K_1(k_{\mu} r)}{K_1(k_{\mu})}
\end{split}
\right.  ,\label{eq:Btheta}
\end{eqnarray}
\begin{eqnarray}
\bB_z=&&-\frac{\ii s}{k}
\left\{
\begin{split}
r< 1,&\;\;\;\lambda_{\sigma}k_{\sigma}\frac{I_0(k_{\sigma} r)}{I_1(k_{\sigma})}+\lambda_{\mu}k_{\mu}\frac{I_0(k_{\mu} r)}{I_1(k_{\mu})}   \\
r> 1,&\;\;\;-\lambda_{\sigma}k_{\sigma}\frac{K_0(k_{\sigma} r)}{K_1(k_{\sigma})}-\lambda_{\mu}k_{\mu}\frac{K_0(k_{\mu} r)}{K_1(k_{\mu})}
\end{split}
\right. , \label{eq:Bz}
\end{eqnarray}
where, again, the coefficient $\exp(\ii k z)$ has been dropped for convenience.
In contrast to $\bJ$, the induction field depends not only on $\sigma$, but also on $\mu$, implying the following remarks.

In the case of identical anisotropic conductivity and permeability, $\sigma=\mu$, as mentioned earlier the dynamo is impossible. From (\ref{eq:ksigma kmu}) and (\ref{eq:A'cont}) we have $k_{\sigma}=k_{\mu}$ and $\lambda_{\sigma}+\lambda_{\mu}=0$, implying that $\bB_r=\bB_z=0$. In that case the induction field $\bB$ is then purely toroidal. This is in agreement with the antidynamo theorem of \citet{Kaiser1994}, after which an invisible dynamo, with a purely toroidal magnetic field, is impossible.

In the limit  $\mu\rightarrow \infty$,  from (\ref{eq:Br}) and (\ref{eq:Btheta}) we have $c \bB_r=-s \bB_{\theta}$, implying that $\bB \cdot \bq=0$.
The projection in the ($r$,$\theta$) plane of the induction field $\bB$ thus describes spiralling trajectories perpendicular to $\bq$.

\begin{figure}
\centering
\includegraphics[scale=0.54]{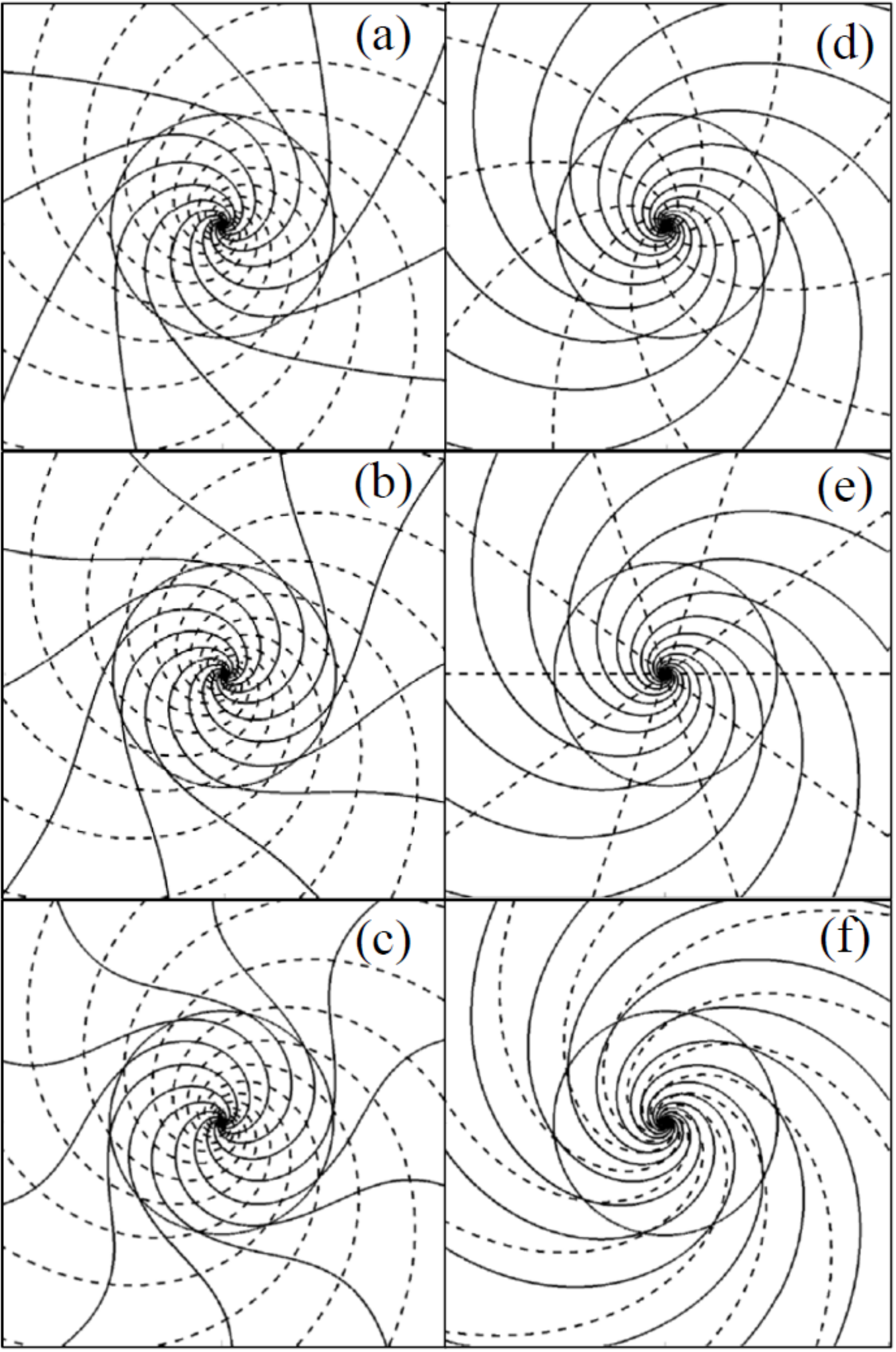}
\caption{Current lines of $\bB$ (solid lines) and $\bJ$ (dashed lines) in the horizontal plane for $\alpha=0.16 \pi$, $k=1.1$, and for (a) $(\sigma,\mu)=(10^5,-0.99)$, (b) $(\sigma,\mu)=(10^5,0)$, (c) $(\sigma,\mu)=(10^5,5)$,
(d) $(\sigma,\mu)=(-0.99,10^5)$, (e) $(\sigma,\mu)=(0,10^5)$, (f) $(\sigma,\mu)=(10,10^5)$.}
\label{fig:spirales2}
\end{figure}

In figure \ref{fig:spirales2} the current lines of $\bB$ and $\bJ$ are plotted in the horizontal plane for different values of $(\sigma,\mu)$. 
In figures \ref{fig:spirales2}a, \ref{fig:spirales2}b and \ref{fig:spirales2}c, $\sigma=10^5$ and $\mu\in\{-0.99;0;5\}$.  The current lines of $\bJ$ are identical because, as previously seen, $\bJ$ does not depend on $\mu$. From \ref{fig:spirales2}a to \ref{fig:spirales2}c, increasing $\mu$ has the effect of distorting the $\bB$ current lines in the outer cylinder, such that the current lines of $\bB$  reach the same  curvature as the current lines of $\bJ$, which eventually is detrimental to dynamo action. In figures \ref{fig:spirales2}d, \ref{fig:spirales2}e and \ref{fig:spirales2}f, $\mu=10^5$ and $\sigma\in\{-0.99;0;10\}$.  From \ref{fig:spirales2}d to \ref{fig:spirales2}f, increasing $\sigma$ has the effect of distorting the $\bJ$ current lines in both inner and outer cylinders, such that the current lines of $\bJ$  reach the same  curvature direction as the current lines of $\bB$, which ultimately is again detrimental to the dynamo action. 
We note that the $\bJ$ current lines in figures \ref{fig:spirales2}a, \ref{fig:spirales2}b and \ref{fig:spirales2}c and the $\bB$ current lines in figures \ref{fig:spirales2}d, \ref{fig:spirales2}e and \ref{fig:spirales2}f are identical. This is because $\sigma \gg 1$ in figures \ref{fig:spirales2}a, \ref{fig:spirales2}b and \ref{fig:spirales2}c, implying $\bJ \cdot \bq \approx 0$,
and $\mu \gg 1$ in figures \ref{fig:spirales2}d, \ref{fig:spirales2}e and \ref{fig:spirales2}f, implying $\bB \cdot \bq \approx 0$.

\section{Dynamo mechanism}
%%%%%%%%%%%%%%
The set of equations (\ref{eq:gamma A}-\ref{eq:gamma B}) can be rewritten in terms of $\bB_r$ and $\bB_{\theta}$ as
\begin{eqnarray}
\gamma \bB_r &=& cs(\sigma -\mu) k^2 \bB_{\theta} - (1+\sigma s^2) D_{\tilde{k}_{\sigma\mu}}(\bB_r) \label{eq:gammaBr}\\
\gamma \bB_{\theta}  &=& cs(\sigma -\mu) D_k(\bB_r) - (1+\mu s^2)D_{\tilde{k}_{\mu\sigma}} (\bB_{\theta})   \label{eq:gammaBteta},
\end{eqnarray}
with 
\begin{equation}
\tilde{k}_{\sigma\mu}=k\left(1+\frac{\mu c^2}{1+\sigma s^2}\right)^{1/2}, \;\;\;\;\; \tilde{k}_{\mu\sigma}=k\left(1+\frac{\sigma c^2}{1+\mu s^2}\right)^{1/2}.
\label{eq:ksigma kmu tilde}
\end{equation}
On the right hand side of each equation (\ref{eq:gammaBr}) and (\ref{eq:gammaBteta}), the first term is a source term for the dynamo effect, while the second term is a decay term.
In (\ref{eq:gammaBr}), resp. (\ref{eq:gammaBteta}), the term $cs(\sigma -\mu) k^2 \bB_{\theta}$, resp. $cs(\sigma -\mu) D_k(\bB_r)$, corresponds to the generation of $\bB_r$ from $\bB_{\theta}$, resp. $\bB_{\theta}$ from $\bB_r$. The differential rotation between the inner and outer cylinders also participates in the generation of $\bB_{\theta}$ from $\bB_r$, through the boundary condition (\ref{eq:Ezcont}). The latter is, however, not sufficient in itself.
Therefore, it is clear why increasing the value of $|\sigma -\mu|$ helps for the dynamo effect, and why the dynamo is impossible for $\sigma=\mu$.
Dynamo action thus occurs through differential rotation conjugated to anisotropic diffusion. 

From the point of view of basic Maxwell and Ohm equations, and in the case of an
isotropic conductivity ($\sigma = 0$) and anisotropic magnetic
permeability ($\mu \neq 0$), dynamo
action can be understood in the following way. Suppose there exists an axisymmetric
magnetic induction disturbance with a non-zero radial component $\bB_r$ at
some height $z$ along the shear zone ($r=1$). Ohm's law (\ref{eq:Ohm}) then drives
two opposite currents in the axial direction $\be_z$ within the rotor and
stator. In a medium of isotropic electrical conductivity, this current
forms closed loops in the meridian planes, as can be seen in
Fig.~6(e). From Amp\`ere's law (\ref{eq:Ampere}), this generates an azimuthal magnetic
field $\bH_{\theta}$. Finally from (\ref{eq:Maxwell1}), 
a radial component of the induction vector $\bB_{r}$ is generated from ${\bH_{\theta}}$
because of
the anisotropic magnetic permeability. Depending on the orientation of the
anisotropic permeability tensor and the direction of the solid rotation, the
generated $\bB_r$ can either reinforce (dynamo
action is possible) or oppose the initial seed of radial magnetic
induction (no dynamo).

\section{Conclusions}
%%%%%%%%%%
For an anisotropic electrical conductivity ($\sigma^{\perp}\neq \sigma^{\parallel}$) conjugated to an anisotropic magnetic permeability ($\mu^{\perp}  \neq \mu^{\parallel}$), we could think that maximizing the ratio $(\sigma^{\perp}\mu^{\perp})  / (\sigma^{\parallel}\mu^{\parallel})$ could help for the dynamo action. This would correspond to minimizing magnetic diffusivity in the perpendicular direction relative to that in the parallel direction.
This is not true for two reasons. First, contrary to the isotropic case, defining an anisotropic magnetic diffusivity is meaningless, because the electrical conductivity and magnetic permeability are now tensors. Second, it has been shown that taking $\sigma^{\perp}  / \sigma^{\parallel} \gg 1$ and $\mu^{\perp}  / \mu^{\parallel} \gg 1$ is in fact highly detrimental to the dynamo effect, these two conditions having the effect of aligning the current lines of respectively the current density $\bJ$ and magnetic induction $\bB$ in the same direction $\bq^{\perp}$.
In contrast, having $\sigma^{\perp}  / \sigma^{\parallel} \gg1$ and $\mu^{\perp}  / \mu^{\parallel}=1$, or $\sigma^{\perp}  / \sigma^{\parallel}  =1$ and $\mu^{\perp}  / \mu^{\parallel}\gg 1$  leads to the same dynamo threshold $|\Omega^c|=14.61$, for $k=1.1$ and $\alpha=0.16\pi$. 

As an application let us consider an experimental demonstration of the dynamo effect based on such conductivity and permeability spiral anisotropy, with differential rotation between two cylinders, as sketched in figure \ref{fig:spirales}. An anisotropic conductivity, resp. permeability, can be manufactured by alternating thin layers of two materials with different conductivities, resp. permeabilities. Although the resulting medium is no longer axisymmetric, our model is still a good approximation of such an experiment.
To realize the first case $\sigma^{\perp}  / \sigma^{\parallel} \gg1$ and $\mu^{\perp}  / \mu^{\parallel}=1$, we can alternate spiral layers of a high electrical conductivity material, e.g. copper, and a material which is electrically insulating, e.g. epoxy resin, both having a relative magnetic permeability equal to unity. To realize the second case $\sigma^{\perp}  / \sigma^{\parallel}  =1$ and $\mu^{\perp}  / \mu^{\parallel}\gg 1$ , we can alternate spiral layers of a high magnetic permeability material, e.g. $\mu$-metal (permalloy), and a material with a relative magnetic permeability equal to unity, e.g. stainless steel, both having approximately the same electrical conductivity. The current lines of $\bB$ and $\bJ$ of these two cases are illustrated  in figure \ref{fig:spirales2}b and \ref{fig:spirales2}e. For the second case, a crucial issue will be to guarantee a good electrical contact between both materials, $\mu$-metal and stainless steel. Indeed, if this is not the case, this would correspond to having $\sigma^{\perp}  / \sigma^{\parallel} \gg 1$ and $\mu^{\perp}  / \mu^{\parallel} \gg 1$ which, again, would be highly detrimental to the dynamo effect.

In the case of an isotropic electrical conductivity, $\sigma^{\perp}  / \sigma^{\parallel}  =1$, and as illustrated in figure \ref{fig:spirales2}e, the azimuthal current density is null, $\bJ_{\theta}=0$ , which is in agreement with the neutral point argument of Cowling. As $\bJ=\nabla \times \bH$, this implies that the circulation of the poloidal component of $\bH$ on a closed current line is zero \citep{Cowling1934}.
However, as shown in (\ref{eq:magnetic field}), in the case of an anisotropic magnetic permeability this does not imply that the poloidal component of $\bB$ is zero. Therefore, although the neutral point argument of Cowling still holds, it does not imply the impossibility of a dynamo effect.

\section*{Acknowledgements}
We thank R. Deguen for fruitful discussions on Earth's inner core anisotropy.

\appendix
%%%%%%%%%
\section{Derivation of $\nabla \times (\bU \times \bB)=0$}
\label{sec: curl of u times B}
%%%%%%%%%%%%%%%%%%%%%%%%%%%%%
Assuming axisymmetry ($\partial_{\theta}=0$), the curl of the cross product of $\bU=r\Omega \be_{\theta}$ and $\bB=\left(B_r, B_{\theta}, B_z \right)$ is given by $\nabla \times (\bU \times \bB)=\left( \partial_z(r\Omega B_z) + \partial_r(r\Omega B_r) \right) \be_{\theta}$. Assuming that $\Omega$ is constant in space and using the solenoidality of $\bB$, $\nabla \cdot \bB=0$, leads to $\nabla \times (\bU \times \bB)=0$.

\section{Derivation of (\ref{eq:gamma A}) and (\ref{eq:gamma B}) }
\label{sec: eq of A and B}
%%%%%%%%%%%%%%%%%%%%%%%%%%%%%%%%%%%
The product of $\lbrack\mu_{ij}\rbrack^{-1}=\begin{pmatrix} 1+\mu c^2 & \mu cs & 0 \\ \mu cs & 1+\mu s^2 &0 \\ 0&0&1 \end{pmatrix}$ given by (\ref{eq:dimensionless permeability}), and the induction field $\bB=\begin{pmatrix} -\ii kA \\ B \\ \frac{1}{r}\partial_r(rA) \end{pmatrix}\exp(\gamma t + \ii kz)$ given by (\ref{eq:magnetic induction}), leads to the magnetic field 
\begin{equation}
\bH=\lbrack\mu_{ij}\rbrack^{-1} \bB=\begin{pmatrix} -\ii k(1+\mu c^2)A+ \mu cs B\\ - \ii k \mu cs A + (1+\mu s^2)B\\ \frac{1}{r}\partial_r(rA) \end{pmatrix},
\label{eq:magnetic field}
\end{equation} 
where, from now, the exponential term is dropped for convenience.
Assuming axisymmetry, the curl of $\bH$ takes the form $\nabla \times \bH = \begin{pmatrix} -\ii k H_{\theta}\\ \ii k H_r - \partial_r H_z\\ \frac{1}{r}\partial_r(rH_{\theta}) \end{pmatrix}$, leading to the current density 
\begin{equation}
\bJ=\nabla \times \bH = \begin{pmatrix} -k^2\mu cs A -\ii k (1+\mu s^2)B \\ D_k(A)+\mu c^2 k^2A+\ii \mu cs k B \\ -\ii \mu csk \frac{1}{r}\partial_r(rA) +(1+\mu s^2)\frac{1}{r}\partial_r(rB) \end{pmatrix}, \label{eq:current density}
\end{equation}
where $D_{\nu}(X)=\nu^2X-\partial_r\left(\frac{1}{r}\partial_r(rX)\right)$.  
The product of $\lbrack \sigma_{ij}\rbrack^{-1}=\begin{pmatrix} 1+\sigma c^2 & \sigma cs & 0 \\ \sigma cs & 1+\sigma s^2 &0 \\ 0&0&1 \end{pmatrix}$ given by (\ref{eq:dimensionless conductivity}), and $\bJ$ given by (\ref{eq:current density}), leads to 
\begin{equation}
\lbrack \sigma_{ij}\rbrack^{-1} \bJ = \begin{pmatrix} -k^2\mu cs A +\sigma cs D_k(A) -\ii k (1+\sigma c^2+\mu s^2)B \\ 
\mu c^2 k^2A+(1+\sigma s^2)D_k(A)-\ii  csk (\sigma - \mu) B \\ -\ii \mu csk \frac{1}{r}\partial_r(rA) +(1+\mu s^2)\frac{1}{r}\partial_r(rB) \end{pmatrix}. \label{eq:sigma current density}
\end{equation}
Taking the curl leads to
$
\nabla \times \lbrack \sigma_{ij}\rbrack^{-1} \bJ =\begin{pmatrix} -\ii k F \\ 
G \\ \frac{1}{r}\partial_r(rF) \end{pmatrix},
$
with 
\begin{eqnarray}
F&=&\mu c^2 k^2 A +(1+\sigma s^2) D_k(A) -\ii k cs (\sigma -\mu)B,\\
G&=&\ii k cs (\sigma -\mu) D_k(A) + \sigma k^2 c^2 B + (1+\mu s^2) D_k(B).\;\;\;\;\;\;\;
\end{eqnarray}
As $\nabla\times(\bU \times \bB)=0$, the induction equation (\ref{eq:induction equation}) is reduced to $\partial_t \bB = -\nabla \times \lbrack \sigma_{ij}\rbrack^{-1} \bJ $, leading to
\begin{eqnarray}
\gamma A &=& -F,\\
\gamma B &=& -G,\\
\frac{\gamma}{r}\partial_r(rA) &=& -\frac{1}{r}\partial_r(rF),
\end{eqnarray}
and then to (\ref{eq:gamma A}) and (\ref{eq:gamma B}).

\section{Derivation of the fourth-order differential equation  (\ref{eq:differential equations}) satisfied by $A$ and $B$ at the dynamo threshold}
\label{sec:derivation of DoD}
%%%%%%%%%%%%%%
Replacing $\gamma=0$ in (\ref{eq:gamma A}) and (\ref{eq:gamma B})  leads to the following system
\begin{eqnarray}
(1+\sigma s^2)D_k(A)  + \mu c^2 k^2 A &=&\ii cs k (\sigma-\mu) B, \label{eq:A2}\\
(1+\mu s^2)D_k(B) + \sigma c^2 k^2 B&=&- \ii csk (\sigma-\mu) D_k(A), \;\;\;\;\;    \label{eq:B2}
\end{eqnarray}
where, again, $D_{\nu}(X)=\nu^2X-\partial_r\left(\frac{1}{r}\partial_r(rX)\right)$. 

It is straightforward to show that
\begin{eqnarray}
(1+\sigma s^2) D_k(X)=(1+\sigma s^2)D_{k_{\sigma}}(X) - \sigma c^2 k^2 X , \;\;\;\;\; \label{eq:sigma X}\\
(1+\mu s^2) D_k(X)=(1+\mu s^2)D_{k_{\mu}}(X) - \mu c^2 k^2 X, \;\;\;\;\; \label{eq:mu X}
\end{eqnarray}
where $k_{\sigma}$ and $k_{\mu}$ are defined in (\ref{eq:ksigma kmu}) and that we rewrite here for convenience
\begin{equation}
k_{\sigma}=k\left(\frac{1+\sigma}{1+\sigma s^2}\right)^{1/2}, \;\;\;\;\; k_{\mu}=k\left(\frac{1+\mu}{1+\mu s^2}\right)^{1/2}.
\nonumber
\end{equation}

Using (\ref{eq:sigma X}) and (\ref{eq:mu X}) in (\ref{eq:A2})  and (\ref{eq:B2}) then leads to
\begin{eqnarray}
(1+\sigma s^2)D_{k_{\sigma}}(A)   &=&ck (\sigma-\mu) (ckA + \ii sB), \label{eq:A3}\\
(1+\mu s^2)D_{k_{\mu}}(B) &=&- ck (\sigma-\mu) (ckB+\ii sD_k(A)). \;\;\;\;\;    \label{eq:B3}
\end{eqnarray}

Then to obtain (\ref{eq:differential equations}) we need to demonstrate that $D_{k_{\mu}}(ckA + \ii sB)=0$ and $D_{k_{\sigma}}(ckB+\ii sD_k(A))=0$.
For that, we rewrite (\ref{eq:A2}) and (\ref{eq:B2}) as
\begin{eqnarray}
D_k(A)  &=& \sigma \left(-s^2 D_k(A)+ \ii cs k B \right)
- \mu \left( c^2 k^2 A + \ii cs k B\right) , \label{eq:A4}\\
D_k(B) &=& - \sigma\left( c^2 k^2 B+ \ii csk D_k(A)\right) 
+ \mu D_k\left( -s^2 B + \ii cs k A\right).    \label{eq:B4}
\end{eqnarray}
Multiplying (\ref{eq:A4}) by $ck$, (\ref{eq:B4}) by $\ii s$, and adding both quantities leads to 
\begin{equation}
(1+\mu s^2) D_k(ckA+\ii sB)= - \mu c^2 k^2 (ckA+\ii sB),
\end{equation}
which, from (\ref{eq:mu X}) with $X=ckA+\ii sB$, is equivalent to 
\begin{equation}
D_{k_{\mu}}(ckA + \ii sB)=0. \label{eq:A5}
\end{equation}
Applying (\ref{eq:A5})  to (\ref{eq:A3}) then leads to
\begin{equation}
\left(D_{k_{\mu}}\circ D_{k_{\sigma}}\right)(A) = 0. \label{eq:DkmoDks}
\end{equation}
Taking $D_k$ of (\ref{eq:A4}) multiplied by $\ii s $ on the one hand, and (\ref{eq:B4}) multiplied by $ck$ on the other hand, and adding both quantities leads to 
\begin{equation}
(1+\sigma s^2) D_k(ckB+\ii s D_k(A))= - \sigma c^2 k^2 (ckB+\ii s D_k(A)),
\end{equation}
which, from (\ref{eq:sigma X}) with $X=ckB+\ii s D_k(A)$, is equivalent to 
\begin{equation}
D_{k_{\sigma}}(ckB + \ii s D_k(A))=0. \label{eq:B5}
\end{equation}
Applying  (\ref{eq:B5}) to (\ref{eq:B3}) leads to
\begin{equation}
\left(D_{k_{\sigma}}\circ D_{k_{\mu}}\right)(B) = 0, \label{eq:DksoDkm}
\end{equation}
which, together with (\ref{eq:DkmoDks}), corresponds to (\ref{eq:differential equations}).
\section{Derivation of $B$, given in (\ref{eq:Bparameters}), at the dynamo threshold}
\label{sec:Derivation de B}
%%%%%%%%%%%%%%%%%%%%%%%%%%%%
Starting from (\ref{eq:Aparameters}), which we rewrite here as
\begin{eqnarray}
A=&&
\left\{
\begin{split}
r< 1,&\;\;\;\;\;\alpha_{\sigma} I_1(k_{\sigma} r)+ \alpha_{\mu} I_1(k_{\mu} r)&  \\
r> 1,&\;\;\;\;\;\beta_{\sigma} K_1(k_{\sigma} r) + \beta_{\mu} K_1(k_{\mu} r)&
\end{split}
\right. , \nonumber
\end{eqnarray}
we will derive $B$ from (\ref{eq:gamma A}), which we write here for $\gamma=0$ as
\begin{equation}
B=\left((1+\sigma s^2)D_k(A)  + \mu c^2 k^2 A\right)/ \left(\ii cs k (\sigma-\mu)\right). \label{eq:B gamma=0}\\
\end{equation}
Using the relations (\ref{eq:sigma X}) and (\ref{eq:mu X}), and knowing that, whatever $\nu$, $D_{\nu}(I_1(k_{\nu} r))=D_{\nu}(K_1(k_{\nu} r))=0$ we find that
\begin{eqnarray}
D_k(A)=
\left\{
\begin{split}
r< 1,&\;\;\; -\frac{\sigma c^2 k^2}{1+\sigma s^2}\alpha_{\sigma} I_1(k_{\sigma} r) -\frac{\mu c^2 k^2}{1+\mu s^2} \alpha_{\mu} I_1(k_{\mu} r)&  \\
r> 1,&\;\;\; -\frac{\sigma c^2 k^2}{1+\sigma s^2}\beta_{\sigma} K_1(k_{\sigma} r) -\frac{\mu c^2 k^2}{1+\mu s^2} \beta_{\mu} K_1(k_{\mu} r)&
\end{split}
\right. . 
\label{eq:Dk(A)}
\end{eqnarray}
Then, replacing in (\ref{eq:B gamma=0}) the expressions of $A$ and $D_k(A)$ given by (\ref{eq:Aparameters}) and (\ref{eq:Dk(A)}) leads to the following expression for $B$, which is also given in (\ref{eq:Bparameters}):
\begin{eqnarray}
B=&&
\left\{
\begin{split}
r< 1,&\;\;\;\frac{\ii ck}{s} \left(\alpha_{\sigma} I_1(k_{\sigma} r)+ \frac{\mu s^2}{1+\mu s^2}\alpha_{\mu} I_1(k_{\mu} r)\right) &  \\
r> 1,&\;\;\;\frac{\ii ck}{s} \left(\beta_{\sigma} K_1(k_{\sigma} r) + \frac{\mu s^2}{1+\mu s^2}\beta_{\mu} K_1(k_{\mu} r) \right)&
\end{split}
\right. . \nonumber
\end{eqnarray}

\section{Derivation of the boundary conditions (\ref{eq:BC1}), (\ref{eq:BC2}) and (\ref{eq:contdA2}) at the dynamo threshold}
\label{sec:derivation of BC}
%%%%%%%%%%%%%%%%%%%%%%%%%%%%%%%%%%%%%%%
The continuity of $A$ and $B$ at $r=1$, taken from their expressions (\ref{eq:Aparameters}) and (\ref{eq:Bparameters}), takes the following form:
\begin{eqnarray}
\alpha_{\sigma} I_1(k_{\sigma})+\alpha_{\mu} I_1(k_{\mu})&=&\beta_{\sigma} K_1(k_{\sigma})+\beta_{\mu} K_1(k_{\mu})\\
\alpha_{\sigma} I_1(k_{\sigma})+\frac{\mu s^2}{1+\mu s^2}\alpha_{\mu} I_1(k_{\mu})&=&\beta_{\sigma} K_1(k_{\sigma})+\frac{\mu s^2}{1+\mu s^2}\beta_{\mu} K_1(k_{\mu})
\end{eqnarray}
leading to (\ref{eq:BC1}) and (\ref{eq:BC2}), which we rewrite here as
\begin{eqnarray}
\alpha_{\sigma} I_1(k_{\sigma})-\beta_{\sigma} K_1(k_{\sigma})&=&0\nonumber\\
\alpha_{\mu} I_1(k_{\mu})-\beta_{\mu} K_1(k_{\mu})&=&0.\nonumber
\end{eqnarray}
To write the continuity of $\partial_r A$ at $r=1$ we first need to calculate the expression of $\partial_r A$ at any $r$.
Using the following relations satisfied whatever $\nu$:
\begin{eqnarray}
\partial_r \left(I_1(\nu r)\right)&=&\nu I_0(\nu r) -\frac{1}{r}I_1(\nu r),\label{eq:diffI1}\\
\partial_r \left(K_1(\nu r)\right)&=&-\nu K_0(\nu r) -\frac{1}{r}K_1(\nu r),\label{eq:diffK1}
\end{eqnarray}
the expression of  $\partial_r A$ is obtained by deriving (\ref{eq:Aparameters}): 
\begin{eqnarray}
\partial_r A=&&
\left\{
\begin{split}
r< 1,&\;\;\;\;\;\alpha_{\sigma} \left(k_{\sigma} I_0(k_{\sigma} r) -\frac{1}{r}I_1(k_{\sigma} r)\right)
+ \alpha_{\mu} \left(k_{\mu} I_0(k_{\mu} r) -\frac{1}{r}I_1(k_{\mu} r)\right)&  \\
r> 1,&\;\;\;\;\;\beta_{\sigma} \left( -k_{\sigma} K_0(k_{\sigma} r) -\frac{1}{r}K_1(k_{\sigma} r) \right)
+ \beta_{\mu} \left( -k_{\mu} K_0(k_{\mu} r) -\frac{1}{r}K_1(k_{\mu} r) \right)&
\end{split}
\right. . \;\;\;\;\;
\label{eq:dAdr}
\end{eqnarray}
Then, the continuity of $\partial_r A$ at $r=1$ leads to
\begin{eqnarray}
&&\alpha_{\sigma} \left(k_{\sigma} I_0(k_{\sigma}) -I_1(k_{\sigma})\right)+ \alpha_{\mu} \left(k_{\mu} I_0(k_{\mu}) -I_1(k_{\mu})\right) = \nonumber\\
&&\beta_{\sigma} \left( -k_{\sigma} K_0(k_{\sigma})- K_1(k_{\sigma}) \right)+ \beta_{\mu} \left( -k_{\mu} K_0(k_{\mu}) -K_1(k_{\mu}) \right).
\label{eq:contdA3}
\end{eqnarray}
Then, taking advantage of (\ref{eq:BC1}) and (\ref{eq:BC2}), (\ref{eq:contdA3}) can be simplified to
\begin{equation}
\alpha_{\sigma}k_{\sigma} I_0(k_{\sigma})+ \alpha_{\mu}k_{\mu} I_0(k_{\mu})+
\beta_{\sigma}k_{\sigma} K_0(k_{\sigma})+ \beta_{\mu}k_{\mu} K_0(k_{\mu})=0, \nonumber
\end{equation}
which is (\ref{eq:contdA2}).

\section{Derivation of the current density $\bJ$ at the dynamo threshold}
\label{sec:derivation of j}
%%%%%%%%%%%%%%%%%%%%%%%%%%%%%%%%%%%%%%%
We rewrite the current density $\bJ$ which is given in (\ref{eq:current density}) as
\begin{equation}
\bJ=\nabla \times \bH = \begin{pmatrix} -\ii k \phi \\ D_k(A)+\mu c^2 k^2A+\ii \mu cs k B \\\frac{1}{r}\partial_r\left(r\phi \right) \end{pmatrix}, \nonumber
\end{equation}
with $\phi = -\ii k\mu cs A+ (1+\mu s^2)B$.
At the dynamo threshold $A$ and $B$ can be replaced by their expressions (\ref{eq:Aparameters}) and (\ref{eq:Bparameters}), leading to
\begin{eqnarray}
\phi=&&
\left\{
\begin{split}
r< 1,&\;\;\;\frac{\ii ck}{s} \alpha_{\sigma} I_1(k_{\sigma} r) &  \\
r> 1,&\;\;\;\frac{\ii ck}{s} \beta_{\sigma} K_1(k_{\sigma} r) &
\end{split}
\right. . 
\end{eqnarray}
Using the relations (\ref{eq:diffI1}) and (\ref{eq:diffK1}) leads to 
\begin{eqnarray}
\frac{1}{r}\partial_r\left(r\phi \right)=&&
\left\{
\begin{split}
r< 1,&\;\;\;\frac{\ii ck}{s} \alpha_{\sigma} k_{\sigma}I_0(k_{\sigma} r) &  \\
r> 1,&\;\;\;-\frac{\ii ck}{s} \beta_{\sigma} k_{\sigma}K_0(k_{\sigma} r) &
\end{split}
\right. . 
\end{eqnarray}
Using (\ref{eq:mu X}), we find that 
\begin{equation}
D_k(A)+\mu c^2 k^2A+\ii \mu cs k B = D_{k_{\mu}}(A)+\frac{\ii \mu csk}{1+\mu s^2}\phi.
\end{equation}
Then from the expression of $A$ given at the threshold by (\ref{eq:Aparameters}), we have
\begin{eqnarray}
D_{k_{\mu}}(A)=&&
\left\{
\begin{split}
r< 1,&\;\;\;\alpha_{\sigma} D_{k_{\mu}}(I_1(k_{\sigma} r)) &  \\
r> 1,&\;\;\;\beta_{\sigma} D_{k_{\mu}}(K_1(k_{\sigma} r)) &
\end{split}
\right. . 
\end{eqnarray}
Combining (\ref{eq:sigma X}) and (\ref{eq:mu X}) we have
\begin{equation}
D_{k_{\mu}}(X)=D_{k_{\sigma}}(X) +c^2k^2\left( \frac{\mu}{1+\mu s^2}- \frac{\sigma}{1+\sigma s^2} \right) X , \;\;\;\;\; \label{eq:sigmamu X}
\end{equation}
implying that
\begin{eqnarray}
D_{k_{\mu}}(A)=&&
\left\{
\begin{split}
r< 1,&\;\;\;\alpha_{\sigma} c^2k^2\left( \frac{\mu}{1+\mu s^2}- \frac{\sigma}{1+\sigma s^2} \right)I_1(k_{\sigma} r) &  \\
r> 1,&\;\;\;\beta_{\sigma} c^2k^2\left( \frac{\mu}{1+\mu s^2}- \frac{\sigma}{1+\sigma s^2} \right)K_1(k_{\sigma} r) &
\end{split}
\right. , 
\end{eqnarray}
where, again, we used the property that, whatever $\nu$, $D_{\nu}(I_1(k_{\nu} r))=D_{\nu}(K_1(k_{\nu} r))=0$.
Therefore we find that
\begin{eqnarray}
D_{k_{\mu}}(A)+\frac{\ii \mu csk}{1+\mu s^2}\phi=&&
\left\{
\begin{split}
r< 1,&\;\;\;- \frac{\sigma c^2k^2}{1+\sigma s^2} \alpha_{\sigma} I_1(k_{\sigma} r) &  \\
r> 1,&\;\;\;- \frac{\sigma c^2k^2}{1+\sigma s^2} \beta_{\sigma} K_1(k_{\sigma} r) &
\end{split}
\right. . 
\end{eqnarray}
Then the current density takes the following form
\begin{equation}
\mbox{for}\;\; r<1,\;\; \bJ=\begin{pmatrix} 
\frac{ck^2}{s} \alpha_{\sigma} I_1(k_{\sigma} r) \\ 
- \frac{\sigma c^2k^2}{1+\sigma s^2} \alpha_{\sigma} I_1(k_{\sigma} r) \\
\frac{\ii ck}{s} \alpha_{\sigma} k_{\sigma}I_0(k_{\sigma} r) \end{pmatrix},
\end{equation}
\begin{equation}
\mbox{for}\;\; r>1,\;\; \bJ=\begin{pmatrix} 
\frac{ck^2}{s} \beta_{\sigma} K_1(k_{\sigma} r) \\ 
- \frac{\sigma c^2k^2}{1+\sigma s^2} \beta_{\sigma} K_1(k_{\sigma} r) \\
-\frac{\ii ck}{s} \beta_{\sigma} k_{\sigma}K_0(k_{\sigma} r) \end{pmatrix}.
\end{equation}
Then, substituting $\alpha_{\sigma}$ and $\beta_{\sigma}$ by their expressions in terms of $\lambda_{\sigma}$, $\alpha_{\sigma}= -\ii s \lambda_{\sigma} /(kI_1(k_{\sigma}))$ and $\beta_{\sigma}= -\ii s \lambda_{\sigma} /(kK_1(k_{\sigma}))$, leads to (\ref{eq:jr}-\ref{eq:jz}).
\section{Derivation of the antisymmetric relation (\ref{eq:antisymmetry Omegac})}
\label{sec:derivation of antisymmetry}
%%%%%%%%%%%%%%%%%%%%%%%%%%%%%%%%%%%%%%%
Let us rewrite the set of equations 
(\ref{eq:gamma A}-\ref{eq:gamma B}) for $\gamma=0$, the boundary conditions (\ref{eq:contA}-\ref{eq:contdA}) and (\ref{eq:Ezcont}):
\begin{eqnarray}
(1+\sigma s^2)D_k(A)  + \mu c^2 k^2 A &=&\ii cs k (\sigma-\mu) B, \label{eq:AA2}\\
(1+\mu s^2)D_k(B) + \sigma c^2 k^2 B&=&- \ii csk (\sigma-\mu) D_k(A), \label{eq:BB2}\\
\left[A\right]^{r=1^+}_{r=1^-}=
\left[B\right]^{r=1^+}_{r=1^-}=
\left[\partial_r A\right]^{r=1^+}_{r=1^-}&=&0,\label{eq:contdAA}\\
(1+\mu s^2)\left[\partial_rB\right]^{r=1^+}_{r=1^-} &=& -\ii k \Omega A(r=1^-), \label{eq:contJzz}
\end{eqnarray}
where $\left[X\right]^{r=1^+}_{r=1^-}=X(r=1^+)-X(r=1^-)$, and (\ref{eq:contJzz}) being derived from (\ref{eq:Ezcont}) using (\ref{eq:current density}). 
The system (\ref{eq:AA2}-\ref{eq:contJzz}) is the complete system of equations leading to the dynamo threshold (\ref{eq:Omegac}). 

Now, defining the new variables $A'$ and $B'$ as
\begin{eqnarray}
A' &=& - \frac{1 + \sigma s^2}{1 + \mu s^2} A, \label{eq:A'}\\
B' &=& B - \ii k \frac{c}{s} \left(\frac{1 + \sigma s^2 + \mu s^2}{1 + \sigma s^2}\right) A, \label{eq:B'}
\end{eqnarray}
and replacing them into (\ref{eq:AA2}-\ref{eq:contJzz}) leads to
\begin{eqnarray}
(1+\mu s^2)D_k(A')  + \sigma c^2 k^2 A' &=&\ii cs k (\mu-\sigma) B', \label{eq:A'}\\
(1+\sigma s^2)D_k(B') + \mu c^2 k^2 B'&=&- \ii csk (\mu-\sigma) D_k(A'), \label{eq:B'}\\
\left[A'\right]^{r=1^+}_{r=1^-}=
\left[B'\right]^{r=1^+}_{r=1^-}=
\left[\partial_r A'\right]^{r=1^+}_{r=1^-}&=&0,\label{eq:contdA'}\\
(1+\sigma s^2)\left[\partial_rB'\right]^{r=1^+}_{r=1^-} &=& \ii k \Omega A'(r=1^-). \label{eq:contJz'}
\end{eqnarray}
It shows that the new variables $A'$ and $B'$ obey to the same equations as $A$ and $B$, in which $\sigma$ and $\mu$ have been changed to $\mu$ and $\sigma$, and $\Omega$ to $-\Omega$.
 
\bibliographystyle{jpp}

\end{document}